\documentclass[10pt,conference]{IEEEtran}

\AtBeginDocument{%
  \providecommand\BibTeX{{%
    \normalfont B\kern-0.5em{\scshape i\kern-0.25em b}\kern-0.8em\TeX}}}

\usepackage{amsmath,amssymb,amsfonts}
\usepackage{algorithmic}
\usepackage{graphicx}
\usepackage{textcomp}
\usepackage{xcolor, colortbl}
\usepackage{booktabs}
\usepackage{multirow}
\usepackage{xspace}
\usepackage{caption}
\usepackage{subcaption}
\usepackage{footnote}
\usepackage{url}
\usepackage{tcolorbox}
\usepackage[para]{footmisc}
\usepackage{array,ragged2e}
\usepackage{multicol}
\usepackage{listings}
\usepackage{balance}
\usepackage{tipa}
\usepackage{rotating}

\definecolor{Gray}{gray}{0.9}

\def\BibTeX{{\rm B\kern-.05em{\sc i\kern-.025em b}\kern-.08em
    T\kern-.1667em\lower.7ex\hbox{E}\kern-.125emX}}
    
\ifCLASSOPTIONcompsoc
  \usepackage[nocompress]{cite}
\else
  \usepackage{cite}
\fi 

\newcommand{\suggsize}{\textsc{SUGG}\textsubscript{size}\xspace}
\newcommand{\suggalt}{\textsc{SUGG}\textsubscript{alt}\xspace}
\newcommand{\experience}{\textsc{EXP}\xspace}
\newcommand{\suggnum}{\textsc{SUGG}\textsubscript{enabled}\xspace}
\newcommand{\suggcor}{\textsc{SUGG}\textsubscript{rel}\xspace}
\newcommand{\frustration}{\textsc{frustration}\xspace}
\newcommand{\actualproductivity}{\textsc{PROD}\textsubscript{a}\xspace}
\newcommand{\percivedproductivity}{\textsc{PROD}\textsubscript{p}\xspace}

\newcommand{\aitool}{\textsc{AIDT}\xspace}
\newcommand{\aitools}{\textsc{AIDTs}\xspace}

\newcommand{\ie}{\emph{i.e.,}\xspace}
\newcommand{\eg}{\emph{e.g.,}\xspace}

\newcommand{\etal}{\emph{et~al.}\xspace}
\newcommand{\secref}[1]{Section~\ref{#1}\xspace}

\newcommand{\tabref}[1]{Table~\ref{#1}\xspace}

\newcommand{\RQ}[1]{RQ$_{#1}$\xspace}

\newboolean{showcomments}
\setboolean{showcomments}{true}

\ifthenelse{\boolean{showcomments}}
  {\newcommand{\nb}[2]{
    \fbox{\bfseries\sffamily\scriptsize#1}
    {\sf\small$\blacktriangleright$\textit{#2}$\blacktriangleleft$}
   }
  }
  {\newcommand{\nb}[2]{}
  }

\newcommand{\REV}[1]{\textcolor{black}{#1}\xspace}

\definecolor{gray50}{gray}{.5}
\definecolor{gray40}{gray}{.6}
\definecolor{gray30}{gray}{.7}
\definecolor{gray20}{gray}{.8}
\definecolor{gray10}{gray}{.9}
\definecolor{gray05}{gray}{.95}

\newlength\Linewidth
\def\findlength{\setlength\Linewidth\linewidth
  \addtolength\Linewidth{-4\fboxrule}
  \addtolength\Linewidth{-3\fboxsep}
}
\newenvironment{examplebox}{\par\begingroup
  \setlength{\fboxsep}{5pt}\findlength
  \setbox0=\vbox\bgroup\noindent
  \hsize=0.95\linewidth
  \begin{minipage}{0.95\linewidth}\normalsize}
  {\end{minipage}\egroup
  \textcolor{gray20}{\fboxsep1.5pt\fbox
    {\fboxsep5pt\colorbox{gray05}{\normalcolor\box0}}}
  \endgroup\par\noindent
  \normalcolor\ignorespacesafterend}

\newenvironment{resultbox}{\vspace{0.2cm}\par\begingroup
  \setlength{\fboxsep}{5pt}\findlength
  \hspace{-0.5cm}
  \vspace{0.1cm}
  \setbox0=\vbox\bgroup\noindent
  \hsize=0.95\linewidth
  \begin{minipage}{0.95\linewidth}\normalsize}
  {\end{minipage}\egroup
  \textcolor{gray20}{\fboxsep1.5pt\fbox
    {\fboxsep5pt\colorbox{white}{\normalcolor\box0}}}
  \endgroup\par\noindent
  \normalcolor\ignorespacesafterend}

\begin{document}

\title{How do Copilot Suggestions Impact \\ Developers' Frustration and Productivity?}

\author{
\IEEEauthorblockN{Emanuela~Guglielmi\IEEEauthorrefmark{1}, 
Venera~Arnoudova\IEEEauthorrefmark{2},
Gabriele~Bavota\IEEEauthorrefmark{3},  Rocco~Oliveto\IEEEauthorrefmark{1}, and Simone~Scalabrino\IEEEauthorrefmark{1}}\\
\IEEEauthorblockA{\IEEEauthorrefmark{1}DEVISER @ University of Molise, Italy}
\IEEEauthorblockA{\IEEEauthorrefmark{2} Washington State University, Pullman, WA}
\IEEEauthorblockA{\IEEEauthorrefmark{3}SEART @  Software Institute, Universit\`a della Svizzera italiana, Switzerland}
}

\maketitle

\IEEEtitleabstractindextext{%
\begin{abstract}
\textit{Context.} AI-based development tools, such as GitHub Copilot, are transforming the software development process by offering real-time code suggestions. These tools promise to improve the productivity by reducing cognitive load and speeding up task completion. Previous exploratory studies, however, show that developers sometimes perceive the automatic suggestions as intrusive. As a result, they feel like their productivity decreased.
\textit{Theory.} We propose two theories on the impact of automatic suggestions on frustration and productivity. First, we hypothesize that experienced developers are frustrated from automatic suggestions (mostly from irrelevant ones), and this also negatively impacts their productivity. Second, we conjecture that novice developers benefit from automatic suggestions, which reduce the frustration caused from being stuck on a technical problem and thus increase their productivity.
\textit{Objective.} We plan to conduct a quasi-experimental study to test our theories. The empirical evidence we will collect will allow us to either \textit{corroborate} or \textit{reject} our theories.
\textit{Method.} We will involve at least 32 developers, both experts and novices. We will ask each of them to complete two software development tasks, one with automatic suggestions enabled and one with them disabled, allowing for within-subject comparisons. We will measure independent and dependent variables by monitoring developers' actions through an IDE plugin and screen recording. Besides, we will collect physiological data through a wearable device. We will use statistical hypothesis tests to study the effects of the treatments (\ie automatic suggestions enabled/disabled) on the outcomes (frustration and productivity).
\end{abstract}

\begin{IEEEkeywords}
    Code Assistant, GitHub Copilot, Productivity, Developer's Frustration
\end{IEEEkeywords}}

%
%
%

\IEEEdisplaynontitleabstractindextext
\IEEEpeerreviewmaketitle


\section{Introduction} \label{sec:intro}
Automation of source code generation has long been a key focus of software engineering research \cite{ciniselli2023source}. Several techniques have emerged to speed up the coding process, including code completion tools that predict subsequent developer actions and retrieval-based approaches that recommend relevant code fragments. In addition, advances in deep learning have made it possible to generate code statements directly from natural language descriptions, bringing us closer to achieving seamless integration between human intent and automatic coding support.
AI-assisted development tools (\aitools), such as GitHub Copilot, are transforming the way software is developed by offering real-time code suggestions \cite{ernst2022ai}. These tools promise to improve developer productivity by reducing cognitive effort and speeding up task completion \cite{mozannar2024reading,vaithilingam2022expectation,zieglerprod2022}. 

While increasing the correctness of such tools is very important to better support developers, their integration with the development environment is fundamental as well. The typical way in which \aitools work is by providing near-real time suggestion directly in the code editor: As soon as a good-enough suggestion is available, they show it to the developer, who can accept, ignore, or reject it.
Such a kind of interaction, however, can be a double-edged sword. 
On the one hand, suggestions can serve as a lifeline for developers who are stuck with a particularly challenging feature to implement or with some APIs they do not know well.
On the other hand, the many suggestions continuously provided by the tool might be perceived as interruptions \cite{ma2024breaking} that require attention, evaluation, and action. As such, paradoxically, they might disrupt developers' cognitive flow, increase their frustration \cite{wermelinger2023using,barke2023grounded,zhang2023practices} (especially if the they are irrelevant or require significant adaptation), and reduce their productivity \cite{wang2024evaluating,prather2023}.


Based on previous empirical results, we conjecture that the impact of automatic suggestions provided by \aitools mostly depends on the developer's experience \cite{vaithilingam2022expectation,peng2023impact}. As training wheels in a bike are fundamental for novices but a burden for experts, so we hypothesize that automated AI suggestions are helpful for novice developers, but not for experts. Specifically, we theorize that novice developers might  benefit from automatic suggestions since receiving proactive help might keep their morale high as compared to the scenario in which they continuously need to ask for it. \REV{As pointed out in the literature increased frustration leads to a negative attitude \cite{prather2023} and that negative emotions reduce productivity \cite{nizamudeen2024investigating}. Our focus is on frustration and productivity.} Instead, we expect expert developers to mostly perceive the automatic suggestions as intrusive, interrupting their cognitive flow and increasing their frustration. Such developers might prefer to keep control over their workflow and work out solutions independently.
Previous work has explored the impact of \aitools on developer productivity and stress in general, and few studies have systematically examined how AI-generated suggestions affect these factors under different conditions \cite{ziegler2022productivity,peng2023impact,prather2023,zieglerprod2022}. While previous studies provide valuable insights into how tools like Copilot influence productivity, to the best of our knowledge, there are not yet any studies that conducted controlled experiments to measure the frustration caused by interruptions from Copilot suggestions.

Based on empirical evidence from previous exploratory studies \cite{vaithilingam2022expectation,peng2023impact,prather2023}, in this paper we propose two theories. First, \textit{expert developers are likely to experience increased frustration and reduced productivity due to automatic suggestions since they may act as interruptions}. Second, \textit{novice developers are likely to benefit from automatic suggestions as they provide valuable support that can reduce frustration and enhance productivity}. We propose the design of a \REV{ quasi-experimental study} we plan to run to prove or disprove such theories. Our research can provide useful feedback to tool makers. Our results will help them better design the UX of \aitools, which should keep developers' needs into account to fully serve their purpose. 

\section{Theory} \label{sec:Related}
We propose two theories that aim at describing how AI assistants might impact the developers' productivity and frustration.

\begin{resultbox}
\textbf{T$_E$.} \textit{Expert} software developers may benefit less from automatic suggestions as compared to novice developers. Indeed, we expect that they \REV{experience them as interruptions} \cite{ma2024breaking} that they can find bothering, especially when the recommendations are wrong. We expect that automatic suggestions are associated with increased frustration and a consequent reduction of productivity.
\end{resultbox}

\begin{resultbox}
\textbf{T$_N$.} \textit{Novice} developers are more likely to benefit from automatic suggestion, since they can provide valuable ideas that can help in finding a solution. They may be more tolerant to wrong recommendations and more rarely experience frustration as compared to expert developers.
\end{resultbox}

We provide below the detailed motivation behind our theories and state the specific hypotheses.

\subsection{T$_E$: Expert Developers and Automatic Suggestions}
Peng \etal \cite{peng2023impact} showed that Copilot's suggestions are perceived as \textit{invasive} by experienced developer. They also showed that the perceived invasiveness is also influenced by specific aspects of the suggestions, such as the \textit{size of suggested code blocks} and the \textit{availability of alternative suggestions}, which affect the ability of developers to choose the most fitting suggestion for their code \cite{prather2023}. 
Stimuli perceived as \textit{invasive} increase the developers' \textit{frustration} since they are a specific kind of \textit{interruption} of the developers' flow. Indeed, Ma \etal \cite{ma2024breaking} showed that interruptions significantly increase stress levels, interrupt cognitive flow, and lengthen task completion times.
We expect that disabling the automatic suggestions results in reduced perceived invasiveness, which does not negatively affect the developers' frustration. Note that developers can freely invoke Copilot. As a result, they can get suggestions only when they actually need them, aligning the tool's functionality with their workflow and reducing cognitive overhead. To summarize, we formulate this first hypothesis:
\begin{resultbox}
 \textbf{H$_E^1$}. Automatic suggestions provided by AI assistants increase expert developers' frustration.
\end{resultbox}

Stress plays a crucial role in cognitive performance, task efficiency, and overall well-being of developers \cite{prather2023,nizamudeen2024investigating}. As previously mentioned, automatic suggestions, while potentially beneficial in some cases, can mostly be perceived as interruptions by expert developers, reducing their focus and imposing a cognitive load on them as they review, accept, or reject them \cite{prather2023,wang2024evaluating}. We expect that such a cognitive load, in conjunction with the increased frustration (H$_E^1$), might result in reduced \textit{actual} and \textit{perceived} productivity. More specifically, we formulate the following hypotheses.
\begin{resultbox}
 \textbf{H$_E^2$}. Automatic suggestions provided by AI assistants reduce the actual productivity of expert developers (computed as percentage of completed task divided by time to complete it).
\end{resultbox}
\begin{resultbox}
 \textbf{H$_E^3$}. Automatic suggestions provided by AI assistants reduce the perceived productivity of expert developers.
\end{resultbox}

\subsection{T$_N$: Novice Developers and Automatic Suggestions}
We expect that the working flow of novice developers is completely different from the one of expert developers. Novice developers might find more often in a \textit{blocked} state in which they require additional knowledge to keep coding (\eg APIs or coding patterns). In this context, we expect that they perceive automatic suggestions as a support during coding activity rather than a nuisance. An automatic suggestion might reduce the active request for help to the AI assistant.  We expect that being in the situation of constantly needing and requesting help might increase the frustration of developers \cite{wang2024evaluating,zhang2023practices,jiang2024analysis}. Note that, as for expert developers, the \textit{size of suggested code blocks} might increase the invasiveness of the suggestion and, thus, increase also the novice developers' frustration \cite{prather2023}. On the other hard, having many \textit{alternative suggestions}, in this case, could increase the frustration of developers since they are required to actively make a choice they might not be able to make \cite{prather2023}. \REV{These are aspects that we assume based on the findings in the literature that support our study, in detail we} formulate this hypothesis:
\begin{resultbox}
 \textbf{H$_N^1$}. Automatic suggestions provided by AI assistants reduce novice developers' frustration.
\end{resultbox}

The frequent call for help we hypothesize for novice developers when automatic suggestions are disabled could lead to increased frustration and cognitive overload during coding tasks. This, in conjunction with the increased frustration (H$_N^1$) might ultimately result lower \textit{productivity}. On the other hand, we expect that novice developers can obtain benefit from the automatic suggestions since they can avoid the effort of either (i) actively calling for help, and (ii) possibly switching context to search for solutions in the browser. We expect that this also makes novice developers generally perceive that they are more productive, even though this mostly depends on the correctness of the suggestions: Incorrect suggestions might make even novice developers feel like they are wasting their time. We formulate the following hypotheses:
\begin{resultbox}
 \textbf{H$_N^2$}. Automatic suggestions provided by AI assistants increase the actual productivity of novice developers (computed as percentage of completed task divided by time to complete it).
\end{resultbox}
\begin{resultbox}
 \textbf{H$_N^3$}. Automatic suggestions provided by AI assistants generally increases the perceived productivity of novice developers (provided that the correctness of the suggestions is sufficiently good).
\end{resultbox}

\vspace{-0.3cm}
\section{Research Question} \label{sec:study}
The \textit{goal} of our study is to corroborate or disprove $T_E$ and $T_N$. 
Our study is steered by the following Research Questions:
\begin{itemize}
 \item \textbf{\RQ{1}:} \textit{Do automatic suggestions increase the frustration for expert developers and reduce it for novice developers?} The first RQ aims at validating the hypotheses H$_E^1$ and H$_N^1$ of our theory.
 
 \item \textbf{\RQ{2}:} \textit{Do automatic suggestions reduce the actual productivity for expert developers and increase it for novice developers?} With this second RQ we want to validate the hypotheses H$_E^2$ and H$_N^2$ of our theory.
 
 \item \textbf{\RQ{3}:} \textit{Do automatic suggestions reduce the perceived productivity for expert developers and increase it for novice developers?} The third RQ aims to validate the hypotheses H$_E^3$ and H$_N^3$ of our theory.
  
\end{itemize}

\section{Study Context}
\label{sec:context}
The context of our study consists of \textit{subjects} (software developers) and \textit{objects} (tasks to be completed).
\subsection{Participants}
We plan to involve at least 32 developers, \REV{with a target of up to 50 participants if feasible, ensuring that} half of which are \textit{experts} and half \textit{novices}. Such a number of participants is in line with similar studies that conducted controlled experiments and interviews, which typically involve between 17 and 32 participants \cite{ma2024breaking,prather2023,tan2024far,barke2023grounded,vaithilingam2022expectation,jiang2024analysis}.
To recruit the novice developers, we will target bachelor and master students in Computer Science who have passed all the basic programming exams (procedural programming and object-oriented programming). \REV{Additionally}, we will recruit expert developers by involving industrial developers, using professional networks and industry collaborations. In both cases, we will adopt convenience sampling, \ie we will invite students at the universities we work in and professionals from industries with which we collaborate. \REV{Given the challenges of recruiting experienced developers, we set 32 as the minimum, ensuring equal representation of experts and novices, but we plan to expand recruitment to 50 participants to strengthen the study’s validity.}

First, we will distribute an invitation survey to the candidates. This survey will help us identify suitable participants for the study based on specific criteria. As in the study conducted by Peng \etal \cite{peng2023impact}, participants' experience will be measured through an initial questionnaire that collects demographic and occupational data, including (i) years of programming experience, (ii) hours of programming per day, (iii) programming language preference, (iv) education level, and (v) employment status.
\REV{Additionally, we will collect information about their prior use of Copilot, including the duration and frequency of use, to ensure a minimum level of familiarity for participation.}
We want to have a clear distinction between \textit{expert} developers and \textit{novice} developers to have a chance to better observe the effects we previously theorized, if present. To do this, we will make sure that the maximum years of programming experience of students (candidate novices) and the minimum of professionals (candidate experts) is of at least 3 years. This will allow us to exclude junior industrial developers, who might have very similar programming experience to master or even bachelor students, and students with previous industrial experience. \REV{To further ensure homogeneity, we will identify and remove outliers in terms of years of experience, preventing extreme cases from affecting the composition of the group.}

\subsection{Coding Tasks}
To assess the ability of participants to address real-world software development challenges, we defined two tasks. Both tasks require that participants introduce a significant evolution to an existing system (maintenance tasks). We do not want to administer tasks in which developers need to write code from scratch because they would spend much time designing the system, which is an activity not aided by \aitools.
The tasks are designed so that they take less than 60 minutes to be completed, on average. We aim to let all developers have a chance to complete all the tasks, so we do not want to artificially interrupt them. However, for practical experimental reasons, we will stop each experiment after 120 minutes (\ie the double of what we expect them to need).
We report an overview of the two tasks we defined below and provide a detailed task description in our online appendix \cite{online_appendix}.

\textbf{Task A: Travel Management System.} This task focuses on enhancing the functionality of an existing project titled ``Travel Management System,'' which manages travel records through a REST API. Developers are tasked with implementing specific features such as searching for travels by destination, retrieving travel statistics, and adding other functionalities like pagination and cancellation endpoints. Each sub-task requires extending the existing API by creating new endpoints, implementing the required business logic, and integrating it with the repository layer for database operations.

\textbf{Task B: BrickBreaker Game.} The task involves maintaining and extending the functionality ``BrickBreaker'', a Java game where players use a paddle to bounce a ball to break bricks. The participants are asked to enhance the game by introducing features such as a menu screen, a high-score system, and new gameplay mechanics. Developers are required to implement a preliminary menu with options to start a new game or view high scores, integrate power-ups into the gameplay, and add features like pause/resume functionality and a game-over screen.

\section{Data Acquisition}
\label{sec:data}
We plan to measure a set of variables, divided in \textit{independent variables} and \textit{dependent variables} (or \textit{outcomes}). We expect the former to have an impact on the latter. In the following, we first report the procedure we will adopt to measure all such variables, and then specify the \textit{factors} (\ie the independent variables that we want to focus on) and the \textit{treatments} (\ie the values of the independent variables we plan to study). \REV{We plan to conduct a quasi-experimental study to analyze how automatic suggestions affect developers' frustration and productivity. While we control for some variables (\eg automatic suggestions enable, task order), other factors, such as the size and relevance of AI-generated suggestions are not directly manipulated but observed and then analyzed.}

\subsubsection{Independent Variables}
The independent variables in our study were defined, as previously reported, based on the literature \cite{prather2023,barke2023grounded,zhang2023practices,liang2024large,tan2024far}. We expect all such variables to affect both the dependent variables we aim to study (\ie frustration and productivity) and report them below.

\textbf{\aitool.}
The first fundamental variable that affects frustration and productivity is the specific \aitool. Different \aitools might provide better/worse suggestions, more/less suggestions, might be more/less invasive, and so on. For this reason, we fix this variable and use a single \aitool for all the developers/tasks. We chose GitHub Copilot since it is one of the most popular ones and it is ready to use in existing IDEs (\eg Visual Studio Code).

\textbf{Developer Experience.}
The \experience variable, in our experimental context, is binary: we have \textit{novices} and \textit{experts}. 

\textbf{Automatic Suggestions Enabled.}
The \suggnum variable measures whether developers enabled or disabled automatic suggestions. Again, this is a binary variable.

\textbf{Suggestion Size.} 
The variable \suggsize represents the size of the code block suggested by the \aitool to the developer. We will record the developers' screen during the tasks to collect the code suggestions provided by Copilot. We will measure the average suggestion size in terms of lines of code (LOCs).

\textbf{Number of Alternative Suggestions.}
The \suggalt variable measures the variety of alternative suggestions provided by Copilot. Again, we will use screen recordings to collect such a piece of information. When developers decide to go through several suggestion alternatives, we will count the number of such alternatives and measure the average throughout the task.

\textbf{Relevance of Suggestions.}
The \suggcor variable evaluates whether the suggestions provided by Copilot is accurate to the developer. We will use the Tako plugin for Visual Studio Code to collect such information. Tako records all the developer's activities in the IDE, including the interactions with Copilot. We will measure the percentage of suggestions accepted and rejected by the developer.

\REV{\textbf{Number of Help Request.} We measure the number of manual GitHub Copilot activations (\ie, explicit requests for artificial intelligence-generated suggestions) using the Tako plugin in Visual Studio Code. Tako records all developer interactions with the IDE, including instances where developers manually invoke Copilot for assistance. We will analyze the frequency of manual activations between the two settings (enabled/disabled). It is hypothesized that the number of manual activations is higher when automatic suggestions are disabled, which could be related to increased frustration.}

\subsubsection{Dependent Variables}
The dependent variables reflect the outcomes we aim to observe to prove or disprove our theories.

\textbf{Frustration.}
We measure the \frustration of the developer through physiological data. We will adopt the Empatica E4 Wristband, a medical-grade wearable device that offers real-time physiological data acquisition. We use such a device since it has been widely used in previous research to collect physiological data \cite{girardi2020recognizing,regalia2019multimodal,sevil2017social,zhao2018emotionsense}. Specifically, we will measure the heart rate variability (HRV) throughout each task. Higher variations in HRV indicate greater frustration during task execution \cite{ma2024breaking}. In particular, we will focus on extracting time-domain features of heart rate variability (HRV) from the collected data. In detail, SDNN is the standard deviation of Inter-Beat Intervals (IBIs) measured in milliseconds, where NN means “normal” beats (\ie removing abnormal or false beats). While, RMSSD calculates the difference between successive inter-beat-intervals (IBI) in milliseconds, squares these values, and takes the root of the mean. Punita \etal \cite{punita1970gender} suggested that SDNN and RMSSD were reduced with increased intensity of frustration, and Sin \etal \cite{sin2016linking} found that individuals with more pronounced affective reactivity to stressors had lower levels of SDNN and RMSSD. We are interested in examining the effect of interruptions on developers’ frustration measures and whether it varies among different types of software engineering tasks and developer experience. To extract the time-domain features, we will use \textit{hrvanalysis} \footnote{\url{https://pypi.org/project/hrv-analysis/}}, a Python module specifically designed for HRV analysis. Based on the study of Ma \etal \cite{ma2024breaking} we will capture the 30-second window before and after each interruption occurred during the evaluation. The 30-second window measurement taken before the interruption (\ie Copilot interaction) will be designated as the baseline for subsequent comparisons. Through establishing a baseline in a pilot study, we will try to assess any changes or deviations in physiological responses resulting from the interruptions.
First, we plan to report the differences of SDNN and RMSSD for measurements before and after each interruption. We will perform paired tests with different types of interruptions and software engineering tasks. An increase in RMSSD indicates a decrease in stress measures.
\REV{Our study specifically focuses on stress induced by interruptions during task execution, which has been shown to contribute to frustration. Yuan \etal \cite{yuan2015unconscious} further predict that heart-rate measures will be positively correlated with the negative emotions experienced during frustration. In our case, we hypothesize that the stress measured through HRV is primarily driven by interruptions caused by Copilot’s suggestions. To examine the effect of these interruptions on developers’ frustration levels, we will analyze whether physiological responses vary across different types of software engineering tasks and developer experience.}

\textbf{Actual Productivity.}
For each task completed by each developer, we will measure the \textit{completion percentage} as the number of sub-tasks correctly completed divided by the total number of sub-tasks \cite{ponzanelli2014prompter, ponzanelli2017supporting}, and \textit{completion time}, measured in seconds passed from the beginning of the task to the moment in which the participants says they have completed it.
We compute \actualproductivity as the \textit{completion percentage} divided by \textit{completion time}. \REV{The more tasks a developer completes, the higher the productivity. The more time developers take to complete the task, the lower the productivity.}

\textbf{Perceived Productivity.}
The \percivedproductivity is, by definition, self-reported by developers. To measure it, we ask participants to evaluate how productive they felt during the task through a post-task survey. 

\subsubsection{Factors and Treatments}
As per our theory, we consider two of the previously-defined independent variables as factors: \experience and \suggnum. Both variables have two possible values. Note, however, that while we can manipulate the latter (we can enable or disable Copilot in a given task), we can not do that for the former (experience depends on the developers). For this reason, we will select participants in a balanced way in terms of experience (as previously reported in \secref{sec:context}) and for each participant we will administer the two natural treatments as for \suggnum, \ie \textit{T} (enabled) and \textit{F} (disabled).
Ideally, in the experiment, we should fix all the other variables we expect to influence the outcome. Note, however, that we can not do that in this context since we can not force Copilot to provide suggestions with specific characteristics (\eg a pre-defined specific size). Since we will administer the subjects with two tasks, we expect that from a similar context in terms of code projects the suggestions provided will not vary much in terms of such variables. However, we will check them \textit{a posteriori} and possible handle them by using models (see \secref{sec:procedure}).

\subsection{Group Design}
\REV{Our study follows an AB/BA crossover design, where each participant performs both tasks under different conditions (Copilot enabled and disabled). This design allows within-subject comparisons, improving statistical efficiency. However, we acknowledge the potential for carryover effects, where the experience from the first condition might influence the second. To mitigate this, we implemented a 15-minute break between tasks and counterbalanced the order of treatments. We will also conduct statistical tests for carryover effects to ensure validity of our findings \cite{vegas2015crossover,madeyski2018effect}.}
Since we can manipulate only a single independent variable while each developer has their own fixed \experience, we need to administer a task with automatic suggestions enabled (\suggnum = \textit{T}) and a task with automatic suggestions disabled (\suggnum = \textit{F}). We report in \tabref{tab:experimental_combinations} each combination (groups). We assign developers to the groups beforehand in a random way but making sure that each group contains the same amount of \textit{expert} and \textit{novice} developers.

\begin{table}
\centering
\begin{tabular}{lll}
\toprule
\textbf{Group}         & \textbf{First Task}                    & \textbf{Second Task}     \\ 
\midrule
\#1                    & \suggnum = \textit{T}, Task A   & \suggnum = \textit{F}, Task B \\ 
\#2                    & \suggnum = \textit{T}, Task B   & \suggnum = \textit{F}, Task A \\ 
\#3                    & \suggnum = \textit{F}, Task A   & \suggnum = \textit{T},  Task B \\ 
\#4                    & \suggnum = \textit{F}, Task B   & \suggnum = \textit{T},  Task A \\ 
\bottomrule
\end{tabular}
\caption{Groups in our experiments with the related setting.}
\label{tab:experimental_combinations}
\end{table}

\section{Protocol and Materials}
\label{sec:execution}
We report the experimental protocol designed to systematically evaluate the effects of GitHub Copilot on developer frustration and productivity and the materials (surveys) we will use for the experiment.

\subsection{Experimental Protocol}
The experimental protocol for this study was designed to ensure consistency and accuracy when performing and processing the collected data. Before the experiment, a reminder e-mail will be sent to each participant scheduled for the next day. On the day of the experiment, we will configure the E4 Empatica Frustration Measurement tool at the computer to monitor data acquisition during the experiment. Each participant will have assigned the setting and task execution modes as described in the previous section.
When the participant arrives, the researchers introduce themselves and clearly explains the objectives and procedures of the experiment.
During the study, participants will wear an Empatica E4 Wristband. The device is introduced, with researchers highlighting capturing physiological data such as heart rate, electrodermal activity (EDA), and acceleration. Participants are reassured about the non-invasive nature of the device and its role in the study.

Participants will be scheduled for a four hours in-person session (two hours for each task). During this session, the participant completes the two tasks in a given order and with Copilot enabled or disabled, based on their group. The experiment begins with a calibration phase for the Empatica E4. Participants are invited to relax and view a soothing video featuring natural scenes, lasting two minutes, following the methodology of Fritz \etal \cite{fritz2014using}. This approach was chosen as it has been demonstrated to effectively return participants' physiological features to a baseline level after approximately one minute.
The connection of the device and the signal quality are verified through the E4 real-time application to ensure accurate data capture. Once calibration is complete, the researchers introduce the coding task at hand and briefly present the design of the existing code from which they need to start coding. After that, the \REV{screen recordings and} coding task starts.

After finishing each task, participants are asked to complete a short, five minute survey to provide feedback on their perceived productivity during the task. To ensure participants are not fatigued, a fifteen minute break is provided between the two tasks. This procedure allows us to evaluate both objective productivity metrics, such as correctness and time, and subjective perceptions of productivity, facilitating a comprehensive analysis of Copilot's impact under different settings.

At the end of the tasks, data recording is stopped and the collected data are processed. \REV{Screen recording analysis will be manually performed in the data analysis phase. Specifically, we will manually review these records to note the suggestions provided by Copilot and analyze how developers handle them, including whether or not they ask for alternative suggestions. Through the recordings, we aim to identify common behavioral patterns, such as reliance on Copilot for specific coding tasks or hesitation in accepting suggestions. In addition, the recordings will serve as a double-check of automatically collected data.}

\subsection{Materials}
The materials of our study are both instruments and surveys. Instruments include the Tako plugin for Visual Studio Code\footnote{\url{https://marketplace.visualstudio.com/items?itemName=codelounge.tako}}, which records developer's activities in the IDE, and the Empatica E4 wristband\footnote{\url{https://www.empatica.com/research/e4/}}, which allows us to measure the frustration.

We will use two surveys: an \textit{invitation survey} (before we enroll participants) and a \textit{post-task survey} (after each task).
We designed such surveys following the indications by Ziegler \etal \cite{zieglerprod2022}, both for the collection of developer information and perceived productivity. We report the details of such surveys in our online appendix \cite{online_appendix}.

\textbf{Invitation Survey.}
The purpose of the invitation survey is to basic information of the candidate participants to select actual participants (experts and novices). We will first collect basic \textit{demographic information} of the candidate participants (\eg name and highest level of education completed). Then we will ask for \textit{experience-related information}, including their \textit{programming proficiency} (\eg preferred programming language and years of experience in it \REV{and average daily programming hours}), and, specifically, their \REV{\textit{familiarity} with (i) Java (since the tasks will be performed in such a programming language), and (ii) the specific technology required by the task}. Finally, we will explicitly ask whether participants have had previous experience with Copilot to understand the impact of the familiarity with such a tool on the outcome.

\textbf{Post-Task Survey.}
The purpose of the post-task survey is to collect information related to the perceived productivity (one of the dependent variables in our experiment).
Participants are asked to rate their level of agreement with various aspects of their experience. These statements explore the developers' concerns about the quality of code produced with Copilot's suggestions and whether they perceive improvements in their productivity, task completion speed, and the quality of the solutions.
We will also ask participants to indicate whether Copilot helped them save time by reducing the need for searching external resources. Finally, the survey examines whether the automatic suggestions are perceived as distracting or beneficial in job satisfaction. \REV{In addition, we ask the developer to indicate the level of perceived stress and frustration.} By capturing these subjective experiences, the survey complements objective performance metrics such as task correctness and completion time.

\section{Data Analysis}
\label{sec:procedure}
To answer \RQ{1}, we will compare the \frustration measured in \textit{expert} and \textit{novice} developers both with automatic suggestions enabled and disabled. 
Frustration is measured at each interruption point, considering a time window that includes both the current interruption and previous interruptions. This means that frustration at a given point in time \(t\) is affected not only by the immediate interruption, but also by previous interruptions within the considered window. For example, frustration at the moment \(t\) may be influenced by previous suggestions at the moment \(t - k\), particularly if those suggestions were incorrect or invasive. \REV{To mitigate potential carryover effects when multiple interactions occur in close succession, we conduct a post-analysis review of overlapping time windows. Specifically, we create a filtered dataset in which data points where the baseline window itself contains an interruption are excluded. This allows us to determine whether residual frustration from prior interruptions influences our results. }

For both populations, we will check whether there is a statistically significant difference in terms of frustration in the two scenarios. To this end, we will use a Wilcoxon signed-rank test \cite{wilcoxon1970critical}. The null hypotheses H$_{E}^0$ and H$_{N}^0$ are that there is no difference in terms of frustration when automatic suggestions are enabled and disabled. 
\REV{Since multiple hypothesis tests are performed, we apply a correction for multiple comparisons using the Bonferroni method \cite{weisstein2004bonferroni}. Instead of the standard $\alpha = 0.05$, we use a corrected significance threshold of $\alpha' = 0.017$ (0.05/3).}
We will also report the effect size using Cliff delta \cite{macbeth2011cliff}.
We will check \textit{a-posteriori} whether the variables we expect to be reasonably fixed, \ie \suggsize, \suggalt, and \suggcor, are actually so (\ie they have a negligible variation among different developers) or not.
If one or more variables have clear variations among developers, we will construct two regression models (one for experts and one for novices) to analyze and quantify the relationships between such variables, the \suggnum variable, and \frustration. Such an analysis will be used to identify the contribution of each independent variable to the dependent outcomes.

We will use an analogous procedure to answer \RQ{2} and \RQ{3}. As for the former, we will use \actualproductivity instead of \frustration. As for the latter, we will use \percivedproductivity instead of \frustration.


\section{Limitations, Challenges, and Mitigations} \label{sec:threats}

The first limitation is related to the choice of the tasks, which may not fully represent coding tasks encountered in real work contexts. To address this limitation, we decided to administer two maintenance tasks in which the participants do not need to write code from scratch, which aims to make the tasks more realistic.
We will use the Empatica E4 wristband to measure participants' heart rate variability (HRV). Motion artifacts or inconsistent skin contact can cause noise in the measurements. Given the task at hand, we expect that participants' motion will be minimal (as so the related artifacts). Besides, a researcher will make sure that the device is correctly worn before starting the task.
One of the main challenges of our planned study is to involve a sufficient number of participants. The planned sample size (\textgreater 32 developers) might not be sufficient to observe the phenomena we theorized even if it existed (type II error). \REV{ However, as we plan to involve 50 participants, we can achieve 80\% power with a 50\% difference. We assumed a standard deviation of 50\% as a reasonable and conservative estimate, given that actual productivity ranges between 0 and 1. With a mean of 0.5, this standard deviation would place 95\% of the values within (0.01, 0.99), which is entirely within the expected range.}

\section{Related Work} \label{sec:related-work}
Based on the goal of our study, we discuss previous work focused on the developers' satisfaction with \aitools and how recommendation systems affect developers' productivity.


\textbf{Impact on Frustration.} Prather \etal \cite{prather2023} show that developers are frustrated by incorrect or overly large suggestions that interrupt their workflow, while Liang \etal \cite{liang2024large} found that frustration occurs when the suggestions do not satisfy functional or non-functional requirements. 
Wang \etal \cite{wang2024investigating} report that frustration is caused both by insufficient help and invasive and premature suggestions not aligned with developers' intentions. 
Vaithilingam \etal \cite{vaithilingam2022expectation} found that Copilot, while providing useful starting points and reducing the need for online research, did not significantly improve task completion times or success rates, and participants often struggled to understand, edit, and debug the generated code.

\textbf{Impact on Productivity.} Sun \etal \cite{sun2023don} proposed an early rejection mechanism using a Transformer-based lightweight estimator, significantly reducing computational costs and unhelpful completions. The results emphasize the importance of optimizing code recommendation systems to improve usability and efficiency.
Peng \etal \cite{peng2023impact} conducted a controlled experiment to assess Copilot's impact on productivity. Developers who used Copilot completed a JavaScript-based task 55.8\% faster than those who did not have access to the tool, demonstrating its potential in speeding up development and helping newcomers.
Ziegler \etal \cite{zieglerprod2022} presented a study to assess developer productivity through the comparison of perceived and objective productivity. Their results show the acceptance rate of code suggestions provided by GitHub Copilot is a significant predictor of developers' perceived productivity.

\textbf{Interruptions.} Finally, previous studies measured the effects of interruptions in the workplace and found that they have negative effects such as burden or physical or emotional difficulty \cite{chen2018voluntary,lavie2010attention,lin2007comparison}. Interruptions can increase stress, frustration, time pressure, and the effort required to complete core cognitive tasks \cite{bailey2001effects,murphy2019predicts}.
In software engineering, interruptions play a crucial role in shaping developers' perceptions of productivity and the quality of their workday \cite{meyer2017work,meyer2019today}.
Ma \etal \cite{ma2024breaking} presented a study to explore how different types of interruption influence productivity and stress during software engineering tasks. Using physiological data from wearable sensors and self-reported stress evaluations, the study revealed that high dominance on screen interruptions significantly increased the time required for code comprehension, and a combination of in-person and on-screen interruptions influenced productivity during code review. 
The study finds that interruptions during code writing are associated with greater stress than other tasks. 

\section{Conclusion} \label{sec:conclusion}
We introduced two theories that aim to predict the impact of automatic suggestions in \aitools on developers' frustration and productivity. To test them, we designed a \REV{quasi-experimental study} and proposed an experimental protocol. The results of our experiment will offer valuable feedback for tool makers to refine the user experience of such tools. \REV{Should our hypotheses be supported by empirical evidence, a potential UX improvement could be suggest that Copilot's behavior should be customized based on the user's programming experience. For example, Copilot could be modified to dynamically determine the frequency of automated suggestions based on the developers' experience, also considering the technologies used in the project at hand. Such adaptations could lead to a more efficient and user-friendly AI-assisted coding experience.}


\balance

\bibliographystyle{IEEEtran}
\bibliography{main}

\end{document}